\documentstyle[11pt,paspconf]{article}
\def\lya{Ly$\alpha$}
\def\gapprox{\;\rlap{\lower 2.5pt
 \hbox{$\sim$}}\raise 1.5pt\hbox{$>$}\;}
\def\gsim{\;\rlap{\lower 2.5pt
 \hbox{$\sim$}}\raise 1.5pt\hbox{$>$}\;}
\def\lsimpprox{\;\rlap{\lower 2.5pt
   \hbox{$\sim$}}\raise 1.5pt\hbox{$<$}\;}
\def\lsim{\;\rlap{\lower 2.5pt
   \hbox{$\sim$}}\raise 1.5pt\hbox{$<$}\;}
\begin{document}

\title{Detection of the First Star Clusters With NGST}
\author{Z. Haiman and A. Loeb}
\affil{Astronomy Department, Harvard University, 60 Garden Street,
Cambridge, MA 02138}

\begin{abstract}
We calculate the observable signatures of the first generation of
stars at high redshift ($5<z<100$). To determine the cosmic
star--formation history, we use an extension of the Press--Schechter
formalism for Cold Dark Matter (CDM) cosmologies that incorporates gas
pressure. We calibrate the fraction of gas converted into stars to be
6\% so as to reproduce the 1\% solar C/H ratio observed in the
intergalactic medium (IGM) at $z=3$.

With this star--formation efficiency, we find that NGST would be able
to image more than $10^4$ star clusters from high redshifts ($z>10$)
within its $4^{\prime}\times4^{\prime}$ field of view.  If stars
occupy a region comparable to the virial radius of the cluster, then
$\sim$1\% of these clusters could be resolved.  We calculate the
expected number--flux relation and angular size distribution for these
early star clusters.  We also describe the reionization of the IGM due
to the first generation of stars, and the consequent damping of the
CMB anisotropies on small angular scales.  This damping could be
detected on $\la 10^\circ$ angular scales by MAP and PLANCK.

\end{abstract}

\keywords{cosmology, structure--formation}

\section{Introduction}

In this contribution, we describe calculations of the signatures of
the first generation of stars from high redshifts ($z>5$).  Our
approach is to combine current data on CDM cosmologies and
star--formation in the most straightforward way.  The standard
hierarchical structure formation models predict the abundance of
virialized clouds as a function of mass and redshift.  The first of
these clouds appear with a low mass ($M\sim10^5{\rm M_\odot}$) at
redshifts as high as $z\sim50$; objects with successively higher
masses assemble later (cf. Haiman, Thoul, \& Loeb and references
therein).  Such objects are the most natural sites for the formation
of the first generation of stars.

Although it is not known how star--formation in these objects
proceeds, a necessary requirement for continued post--virialization
collapse and fragmentation is that the gas can cool efficiently.  In
the metal--poor primordial gas, the only coolants available are
neutral atomic hydrogen (H) and molecular hydrogen (${\rm
H_2}$). However, it has been found (Haiman, Rees, \& Loeb 1996, 1997;
Gnedin \& Ostriker 1997) that the ${\rm H_2}$ molecules are fragile,
and are photodissociated throughout the universe.  Accordingly, we
assume that clouds form stars if and only if they are massive enough
to cool via atomic H transition lines ($T\gsim10^4$K).

The total number of stars formed can be fixed using the recent
observations (Songaila \& Cowie 1996, Tytler et al. 1995) of the near
universal carbon to hydrogen (C/H) ratio in Lyman--$\alpha$ forest
clouds at redshifts as high as $z=3$.  These observations yield the
value C/H $\sim$ 1\% solar for systems with a large range of
column--densities.  The fact that (i) the scatter in the metalicity
from one system to another is only an order of magnitude, and (ii) the
overdensity of the lowest column density system is only a few (i.e.,
too tenuous to form its own stars) supports the hypothesis of uniform
metal enrichment by an early generation of stars.

Based on this star--formation efficiency, the redshift evolution of
the abundance of star--forming clouds, and a detailed composite model
spectrum for a low--metalicity stellar population with the local Scalo
(1986) initial mass function (IMF), we derive the expected
number--flux relation and angular size distribution of these star
clusters.  We find that NGST would be able to image more than $10^4$
of these star clusters at $z>10$ within its
$4^{\prime}\times4^{\prime}$ field of view, with $\sim$1\% of these
clusters possibly resolved.

In addition, we show that the pre--galactic population of stars
reionize the universe by a redshift of $z=10-20$.  The consequent
damping of microwave anisotropies on small angular scales is
$\sim10$\%, detectable on $\la 10^\circ$ angular scales by MAP and
PLANCK\footnote{See the homepages for the MAP
(http://map.gsfc.nasa.gov) and PLANCK
(http://astro. estec.esa.nl/SA-general/Projects/Cobras/cobras.html)
experiments.}.  The pre--galactic low--mass stars could also account
for some of the microlensing events observed in the halo of the Milky
Way (Alcock et al. 1997), and could also be detected in the future
through their lensing of distant quasars (Gould 1995).

\section{Description of Model}

To quantify the observational signatures of the first stars in CDM
cosmologies, we use a simple semi--analytic approach that is
complimentary to more detailed, but computationally expensive and
therefore less versatile, 3--D numerical simulations (e.g. Gnedin \&
Ostriker 1997).  The main ingredients of our model are:

\medskip
\noindent
\underline{\em The Collapsed Fraction of Baryons.}  We use the
Press--Schechter formalism to find the abundance and mass distribution
of virialized dark matter halos.  Since initially most of the
collapsed baryons are in low--mass systems near the Jeans mass, the
pressure of the baryons has a significant effect on the collapsed
fraction.  Effectively, the collapse of the baryons is delayed
relative to the dark matter in these low--mass objects (Haiman, Thoul
\& Loeb 1996).  We obtained the exact collapse redshifts of
spherically symmetric perturbations by following the motion of both
the baryonic and the dark matter shells with a one dimensional
hydrodynamics code (Haiman \& Loeb 1997).  We find that because of
shell--crossing with the cold dark matter, baryonic objects with
masses $10^{2-3}{\rm M_{\odot}}$, well below the linear--regime Jeans
mass, are able to collapse by $z$$\sim$$10$.

Another important effect is the feedback on the collapsed fraction
from the photoionization of the IGM.  As the first stars form, they
ionize a fraction $F_{\rm HII}$ of the gas in the universe.  In these
regions condensation of objects is strongly suppressed; to be
conservative we assume that only a fraction ($1-F_{\rm HII}$) of the
gas participates in forming new virialized objects.

\medskip
\noindent
\underline{\em Star Formation.}  
We calibrate the fraction $f_{\rm star}$ of the condensed gas
converted into stars in each virialized cloud using the inferred C/H
ratio in the \lya~absorption forest. We use tabulated $^{12}$C yields
of stars with various masses, and consider three different initial
mass functions (IMFs).  The uncertainty in the total carbon production
is a factor of $\sim$10; a factor of $\sim$3 is from the uncertainty
in the carbon yields of $3-8{\rm M_\odot}$ stars due to the unknown
extent of hot bottom burning (Renzini \& Voli 1981), and another
factor of $\sim$3 is due to the difference between the Scalo and
Miller--Scalo (1979) IMFs.  To be conservative, we assume inefficient
hot bottom burning, i.e. maximum carbon yields.  Under these
assumptions, we find $f_{\rm star}=13\%$.
\footnote{Since the collapsed fraction at $z=3$ is $\sim$50\%, the 
fraction of all baryons in stars is $\sim$6\%.  A factor of $\sim$3
is included in this number due to the average time required to produce 
carbon inside the stars; i.e. only a third of the total stellar carbon 
is produced by $z=3$.}

We also include a negative feedback on star--formation due to the
photodissociation of molecular hydrogen by photons with energies in
the range $11.2$--$13.6$ eV.  These photons are not absorbed by
neutral H and travel freely across the IGM. We find (Haiman, Rees \&
Loeb 1997) that the masses required to self-shield ${\rm H_2}$ against
photodissociation by the Solomon process (cf. Field, Somerville \&
Dressler 1966) are exceedingly high ($M>10^{20}{\rm M_\odot}$ for
a spherical object with an overdensity of 200 at $z=25$).  As a
result, soon after the appearance of the first few stars, ${\rm H_2}$
is universally destroyed, and molecular cooling is suppressed even
inside dense objects.  Due to the lack of any other cooling agent in
the metal--poor primordial gas, the bulk of the pre--galactic stars
must form via atomic line cooling inside massive clouds with virial
temperatures of at least $\sim 10^4$K, or $M_{\rm tot}\gsim10^8
[(1+z)/10]^{-3/2} {\rm M_{\odot}}$.  We therefore allow
star--formation only inside clouds with at least this mass.

\medskip
\noindent
\underline{\em Propagation of Ionization Fronts.} 
To determine the ionization history of the universe, we need to follow
the evolution of the ionization (Str\"omgren) front around each star
cluster.  The composite spectrum of radiation which emerges from each
star cluster is determined by the stellar IMF and the recombination
rate inside the cluster. We follow the time-dependent spectrum of a
star of a given mass based on standard spectral atlases (Kurucz 1993)
and the evolution of the star on the H--R diagram as prescribed by
theoretical evolutionary tracks (Schaller et al. 1992).

To calculate the fraction of the ionizing photons lost to
recombinations inside their parent cloud, we adopt the equilibrium
$1/r^2$ density profile of gas inside each cloud according to our
spherically--symmetric simulations (Haiman, Thoul \& Loeb 1996).  We
assume that stars are distributed with the same $1/r^2$ profile across
the cloud, and obtain the number of recombinations under the
assumption of ionization equilibrium.  We then use the time--dependent
composite luminosity of each star--forming region to calculate the
propagation of a spherical ionization front into the surrounding
homogeneous IGM.  The ionized fraction of the universe is given by the
volume--filling factor $F_{\rm HII}$ of the ionized bubbles, and the
universe is reionized when these bubbles overlap so that $F_{\rm
HII}=1$.

\section{Predictions for NGST}

Figure~1 shows the predicted number of high--redshift star--clusters
at $z>10$ and $z>5$, per logarithmic flux interval, in the wavelength
range of 1--3.5$\mu$m.  The number of clusters that could be probed by
future space telescopes depends on their sensitivity; the vertical
dashed lines show the proposed imaging thresholds of the Space
Infrared Telescope Facility (SIRTF) and the Next Generation Space
Telescope (NGST).  NGST would be able to image $\gsim10^4$ star
clusters at its flux limit from high redshifts ($z>10$) within its
$4^{\prime}\times4^{\prime}$ field of view.

\begin{figure}[t]
\includegraphics{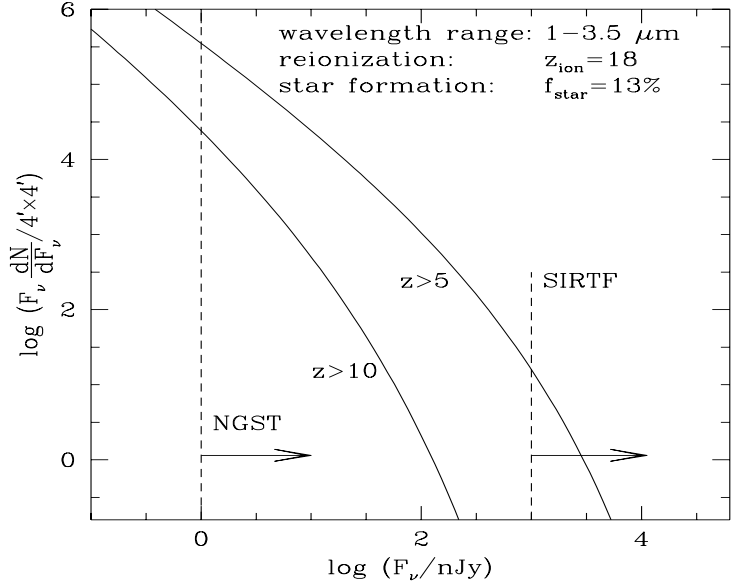}
\vspace{3in}
\caption{Predicted number counts in the 1--3.5$\mu$m range.}
\label{fig-1}
\end{figure}

\begin{figure}[t]
\includegraphics{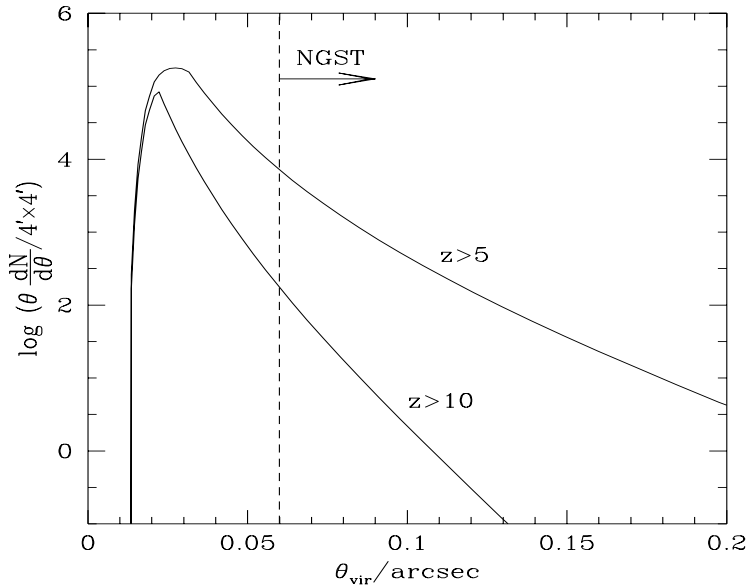}
\vspace{3in}
\caption{Predicted distribution of virial angular diameters.}
\label{fig-2}
\end{figure}

Figure~2 shows the predicted angular diameter ($\theta_{\rm vir}$)
distribution of the star clusters.  $\theta_{\rm vir}$ is taken to be
twice the angle subtended by the virial radius $r_{\rm vir}$ of each
star--cluster; $r_{\rm vir}$ is calculated based on a spherically
symmetric $1/r^2$ density profile, with a factor of 50 overdensity at
the surface $r_{\rm vir}$ of the cloud, assuming $\Omega_{\rm b}=0.05$
(based on the 1--D simulations).  Note that stars in present--day
galaxies form in a much smaller central region; however, the early
star clusters may behave differently and form stars at their
outskirts, since they are denser, and we have imposed the requirement
of efficient atomic line cooling throughout the cloud.  With its
proposed 0.06\H{} angular resolution, NGST could resolve $\sim$100
virialized star--clusters at $z>10$, and $\sim10^4$ clusters at $z>5$
per field of view.

\section{Reionization}

\medskip
\noindent

Table 1 summarizes the reionization redshifts and resulting electron
scattering optical depths that we obtain in our models for a range of
parameters.  We varied the cosmological power spectrum
($\sigma_{8h^{-1}}$, $n$), the baryon density ($\Omega_{\rm b}$), the
star formation efficiency ($f_{\rm star}$), the escape fraction of
ionizing photons ($f_{\rm esc}$), the IMF, and whether or not the
negative feedback due to ${\rm H_2}$ is included.  For almost the
entire range of parameters, the universe is reionized by a redshift
$\gsim 10$.  The only exception occurs when the IMF is strongly tilted
towards low--mass stars.  We considered an unconventional tilt of this
kind by adding a constant 1.7 to the power--law index, while keeping
the IMF fixed at $M=4{\rm M_\odot}$.  With this tilt, the increased
number of low--mass stars could account for the observed microlensing
events.  In this case, however, reionization is strongly suppressed
due to the absence of massive stars which ordinarily dominate the
ionizing flux.

The redshift evolution of the ionized fraction $F_{\rm HII}$ can be
directly converted into the optical depth $\tau$ to electron
scattering to high redshifts. For the range of parameters in Table~1,
we find $\tau\sim 0.05$--$0.1$.  The corresponding damping factor for
the CMB anisotropies (Hu \& White 1997) is $\sim 5$--$10\%$.  Such a
damping could be detected by the MAP and PLANCK satellites, especially
if the polarization of the CMB is measured (Zaldarriaga 1996).

\begin{table}
\caption{Reionization redshift and electron scattering
optical depth for a range of parameters.}
\begin{center}\scriptsize
\begin{tabular}{|c||c|c|c|c|}
\hline
Parameter & Standard Model  & Range Considered & Reionization Redshift & Optical
 Depth \\
\hline
\hline  
$\sigma_{8h^{-1}}$   & 0.67             & 0.67--1.0  & 18--22 & 0.07--0.11 \\
\hline  
$n$                  & 1.0              & 0.8--1.0   & 13--18 & 0.04--0.07 \\
\hline
$\Omega_{\rm b}$     & 0.05             & 0.01--0.1  & 17--19 & 0.02--0.13 \\
\hline
$f_{\rm star}$       & 13\%              & 1\%--40\%  & 12--24 & 0.05--0.09 \\
\hline
$f_{\rm esc}$        & $f_{\rm esc}(z)$ & 3\%--100\% & 11--18 & 0.05--0.07 \\
\hline
IMF tilt ($\beta$)   & 0                & 0--1.69    & 18--1  & 0.01--0.07 \\
\hline
${\rm H_2}$ feedback & yes              & yes/no     & 18--20 & 0.07--0.11 \\
\hline  
\end{tabular}
\end{center}
\end{table}

\acknowledgments
We thank Myungshin Im and Eliot Quataert for useful discussions.

\end{document}